\documentclass[12pt,epsf]{article}
\usepackage{amsmath}
   \usepackage{amscd}
   \usepackage{amsfonts}
   \usepackage{amssymb}
   \usepackage{amsthm}
   \usepackage{epsfig}

   \def\appendix#1{
     \addtocounter{section}{1}
    \setcounter{equation}{0}
     \renewcommand{\thesection}{\Alph{section}}
    \section*{Appendix \thesection\protect\indent \parbox[t]{11.715cm} {#1}}
     \addcontentsline{toc}{section}{Appendix \thesection\ \ \ #1}
     }

   \newcommand{\newsection}{
   \setcounter{equation}{0}
   \section}

   \newcommand{\eq}[1]{\begin{equation} #1 \end{equation}}
   \newcommand{\ar}[1]{\begin{eqnarray} #1 \end{eqnarray}}
   \newcommand{\tr}{\mathop{\mathrm{tr}}\nolimits}
   \newcommand{\Tr}{\mathop{\mathrm{Tr}}\nolimits}

   \def\const{{\rm const}}
   \def\e{{\,\rm e}\,}
   \def\d{\partial}
   \def\D{\delta}
   \def\dd{^{\dagger}}
   \newcommand{\br}[1]{\left( #1 \right)}
   \newcommand{\vev}[1]{\left\langle #1 \right\rangle}
   \newcommand{\rf}[1]{(\ref{#1})}
   \newcommand{\non}{\nonumber \\*}
   \def\ll2{\ll}
   \newcommand{\ket}[1]{\left| #1 \right\rangle}

   \def\N{${\cal N}=4$ }
   
   \def\d0{L}
   \def\La{\Lambda}
   \def\G{\Gamma}
   \def\H{{\cal H}}

\newcommand{\eps}{\epsilon}

\newcommand{\OO}{{\cal O}}

\newcommand{\NN}{{\cal N}}

\newcommand{\be}{\begin{equation}}
\newcommand{\ee}{\end{equation}}
\newcommand{\ben}{\begin{eqnarray}\displaystyle}
\newcommand{\een}{\end{eqnarray}}
\newcommand{\refb}[1]{(\ref{#1})}

\newcommand{\al}{\alpha}

\newcommand{\nt}{\widetilde n}
\newcommand{\ut}{\widetilde u}
\newcommand{\s}{\sigma}

\newcommand{\bL}{\overline\Lambda}

\newcommand{\sectiono}[1]{\section{#1}\setcounter{equation}{0}}

\begin{document}

{}~ \hfill\vbox{\hbox{hep-th/0212208}
\hbox{UUITP-17/02}
\hbox{ITEP-TH-73/02}
}\break

\vskip 1.cm

\centerline{\large \bf The Bethe-Ansatz for $\NN=4$ Super Yang-Mills}
\vspace*{1.5ex}

\vspace*{4.0ex}

\centerline{\large \rm J. A. Minahan\footnote{
joseph.minahan@teorfys.uu.se}  and K. Zarembo\footnote{
also at ITEP, Moscow, Russia.  konstantin.zarembo@teorfys.uu.se}}
\vspace*{2.5ex}
\centerline{\large \it Department of Theoretical Physics}
\centerline{\large \it Box 803, SE-751 08 Uppsala, Sweden}
\vspace*{3.0ex}

\vspace*{4.5ex}
\medskip
\bigskip\bigskip
\centerline {\bf Abstract}

We derive the one loop mixing matrix for anomalous dimensions in $\NN=4$
Super Yang-Mills.  We show that this matrix can be identified with the
Hamiltonian of an integrable $SO(6)$ spin chain with vector sites.  We
then use the Bethe ansatz to find a recipe for computing anomalous
dimensions for a wide range of operators.  We give exact results for
BMN operators with two impurities and results up to and including first order
$1/J$ corrections for BMN operators with many impurities.  We then use
a result of Reshetikhin's to find the exact one-loop anomalous dimension
for an $SO(6)$ singlet in the limit of large bare dimension.  We also
show that this last anomalous dimension  is proportional
to the square root of the string level in the weak coupling limit.

\bigskip

\vfill \eject
\baselineskip=17pt

\section{Introduction}

   \def\ads{$AdS_5\times S^5$ }

   One of the main results of the AdS/CFT correspondence is that
   individual string states are mapped to local gauge-invariant operators
   in a dual field theory  \cite{9711200,9802109,9802150}. 
But even in the most well understood case
of $\NN=4$ Super Yang-Mills (SYM) this mapping is only known for a
small subset of the operators.  The difficulty in making this mapping
explicit is two-fold: i) String quantization on an \ads background is still
unsolved.  ii) The spectrum of gauge invariant operators is somewhat difficult
to compute.

Previously, it was known that the
chiral primaries in the gauge theory are dual to the string states that survive
the supergravity limit.
More recently it was realized how to go beyond the chiral primaries
by considering operators with large $R$-charges, $J$ \cite{0202021}.  
On the string side
this corresponds to semiclassical states with large
   angular momentum on the $S^5$. For such states,
the \ads geometry essentially reduces via a Penrose limit
to a plane wave geometry \cite{0202111,0201081,0110242}.  
String theory on the plane wave background
is solvable  \cite{0112044,0202109} and an identification can be made between 
the string states
and the gauge invariant operators.
The string quantization on the plane wave is simple enough, at least in the light-cone  gauge, where 
   all string states are generated by an infinite set of creation operators 
similar to those in
    flat space \cite{0112044,0202109}.
    
Amazingly, the operators dual to each of the eigenstates of
   the light-cone string Hamiltonian can be identified. These (BMN) operators are \cite{0202021}:
   \ar{
   \ket{0;J}&\Longleftrightarrow&\tr Z^J, \non
   a^i_{0}{}\dd\ket{0;J}&\Longleftrightarrow&\tr \Phi_i Z^J,\non
   a^i_{n}{}\dd a^j_{-n}{}\dd\ket{0;J}&\Longleftrightarrow&\sum_l
   \e^{2\pi iln/J} \tr \Phi_i Z^l \Phi_j Z^{J-l},\nonumber
   }
   and so on. Here, $\Phi_i$, $i=1,\ldots,6$ are the six scalar fields of \N SYM in the
   adjoint 
   representation of $SU(N)$, and $Z=\Phi_1+i\Phi_2$. The BMN operators have
   charge
   $J$ under the generator of the R symmetry group, which rotates $\Phi_1$ 
into $\Phi_2$.
   On the string side, $J$ is essentially the length of the string on the
   light-cone. 
   The chain of $Z$s can be regarded as a field-theory realization of the string,
   which
   emerges as a compound of $J$ constituents, much in the 
   spirit of the string-bit models \cite{giles}. String excitations are represented by
   impurities
   inserted in the chain \cite{0202021}.

 String theory makes a prediction for 
the anomalous dimensions of the BMN operators at 
   any value of the Yang-Mills coupling in the large-$N$ limit,
by equating the mass of a string state with the full dimension of an
operator \cite{0202021}. 
This prediction can
   be   verified by explicit perturbative 
calculations \cite{0202021,0205066,0206079}.  Furthermore,
one can incorporate stringy corrections in the effective string coupling
$J^2/N$ and compare the results of the string calculations with
the gauge theory computations \cite{0205033,0205048,0205089},
\cite{0208041}--\cite{0212118}.

Inverting the logic we can
   say that by resolving the mixing of operators with two or more
   impurities,
   order by order in perturbation theory,
   one can reconstruct the string spectrum
by computing the anomalous dimensions of operators.
   % which 
   %diagonalize the mixing matrix. 
   We will follow this logic in an attempt to better
   understand the operator/string   correpondence for a wider class of 
    string states, including those states
that are outside of the semiclassical regime \cite{0204051}. 
%Such states may or may not correspond to
%   classical string solutions. 
%A semiclassical state is in general described 
%   not by a single classical trajectory but  by many
%   such trajectories. Indeed, any string state with the wave function of the
%   string's center of mass given by an $S^5$ spherical harmonic of
%   sufficiently
%   high order is semiclassical. The BMN vacuum $\ket{0;J}$ corresponds to the
%   spherical function $(\sin\psi)^J\e^{iJ\varphi}$, where $\psi$ and $\varphi$ are
%   asimuthal and polar angles on $S^5$. The wave function of the BMN vacuum
%    is strongly localized
%   near the equator ($\psi=\pi/2$) at large $J$ and is described by a single
%   classical
%   solution, a relativistic point-like string rotating along a great circle of $S^5$. 
%However, there are many more spherical functions in large representations
 %  of $SO(6)$, which are rapidly oscillating, but delocalized functions of  the
 %  $ S^5$ angles.
 %  They correspond to more general scalar operators in the SYM theory.
These states would correspond to operators made of scalar
fields and with high engineering
dimension but in low representations of $SO(6)$.  

   %In a hope to learn more about strings in $AdS_5\times S^5$,
In this paper
   we will consider mixing of generic scalar operators 
   $\tr\Phi_{i_1} \ldots \Phi_{ i_{L}} $ to  one-loop order 
   in  SYM perturbation theory.
   The problem appears difficult, not only because the number of operators
    grows rapidly with $L$ (roughly as $6^L$), but also because
the operators mix in a way
   which at first sight seems hopelessly entangled. 
However, we are able to make progress in solving
   this problem by establishing an equivalence of the mixing matrix
   with the Hamiltonian of a certain integrable spin chain. This equivalence 
   will allow
   us to use powerful  techniques of the
   algebraic Bethe ansatz \cite{Bethe,Faddeev:ft,Resh1,Resh2}
%, as developed by Reshetikhin \cite{Resh1,Resh2},
to diagonalize the mixing matrix.  In particular, we will find that
the problem of finding the one-loop anomalous dimensions comes down
to solving a set of Bethe equations.

Among the results contained in this paper, we are able to reproduce easily
recent
results \cite{0211032}  for the one loop anomalous dimensions
of BMN operators with two 
impurities.
We then extend these results to 
a large class of BMN operators with more than two impurities.  
We are able to  identify BMN states with the corresponding Bethe states, where among other things, we show that a ``bound state''
containing $M$ Bethe roots extending into the complex plane
 corresponds to having string states with $M$ identical 
oscillators. We also give a
 recipe for finding $1/J$ corrections to the anomalous
dimensions including the  
 explicit results for the first order  corrections. These corrections
are important since they correspond to curvature
corrections away from the plane-wave
background in the full \ads  \cite{0204226,sch,0208010}.

We then go beyond the BMN limit in two explicit examples.  The first example
corresponds to an $SO(6)$ singlet made up of $L$ scalar fields.  In the
large $L$ limit this can be solved explicitly \cite{Resh3}, 
and in fact corresponds
to the operator   made up only of scalars 
that has the largest anomalous dimension for bare dimension
$L$.  We find the anomalous dimension and 
demonstrate that it is linear in $L$.  We also argue that the string level
behaves roughly as $L^2$, so the full dimension of the operator is proportional
to the square-root of the level, a result that follows from AdS string theory
 in strong coupling for
generic operators \cite{9802109}.  The second example is the direct analog of the Heisenberg
anti-ferromagnet, where we also find the anomalous dimension and show that it
is linear in $L$.  We also show how to put in ``holes'' on these states
and explicitly compute the changes in the anomalous dimensions coming from
the holes. The holes can be either $SO(6)$ vectors or  one of the $SO(6)$
spinors.

   Integrable structures have previously appeared in
   string theory for  generalizations of the plane-wave 
background \cite{0207284,0208114,0211257}. 
%In this case, by modifying the five-form field strength, one ends up
%with a supersymmetric sine-Gordon model on the world-sheet.
It is not clear if there is a relation between this integrability and the
integrability discussed in this paper. But it might 
indicate that the integrability  encountered here
   is not accidental but is a manifestation of some general principle
   yet to be found.
We should also mention that integrable spin chains arise in perturbative
analysis of Regge scattering in large-$N$ QCD and Bethe ansatz techniques
were extensively applied there \cite{9311037,9404173,9501232,9902375,0204183}.

In section 2 we derive the one loop mixing matrix for all scalar operators. 
In section 3 we use this matrix to compute the anomalous dimensions for a few
simple examples.  In section 4 we give a brief review of Reshetikhin's
proof of integrability for the $SO(6)$ vector chain and his solution for
the eigenvalues
of the transfer matrix in terms of Bethe roots,  along with the Bethe equations
the roots must satisfy.  In section 5 we use the results from the previous
section to compute the anomalous dimensions for two impurities to all orders
in $1/J$ and for many impurities to first order in $1/J$.  In section 6
we describe solutions to the Bethe equations \cite{Resh3}
which correspond to operators  outside the BMN limit.
We compute the anomalous dimensions for these operators and for nearby operators.   In section 7 we give our conclusions.

   \section{Anomalous dimensions from the spin system}

   We will study one-loop renormalization for all scalar operators
   without derivatives: %in the
   %SYM theory:
   \eq{\label{natur}
   \OO[\psi]=\psi^{i_1\ldots i_{\d0}}\tr\Phi_{i_1} \ldots \Phi_{ i_{\d0}}\,.
   }
   Many interesting operators in \NN=4 SYM, notably
   chiral primary and BMN operators, belong to this class. % of operators.
   In general, the scalar operators \rf{natur}
    mix under renormalization. There is a distinguished basis, in which operators
   are multiplicatively renormalizable. It is important that up to possible
   degeneracies,
   rotations to this basis will diagonalize the
    two-point correlation functions. 
   As far as one-loop renormalization is concerned, the
   scalar operators will mix only among themselves.
   Mixing with other operators should occur at  higher orders
   in perturbation theory.

   Renormalized operators  in general are linear combinations
   of bare operators. If we choose the particular operator basis,
   \eq{
   \OO^A_{\rm ren}=Z^A_{\hphantom{A}B}\OO^B,
   }  
   %Correlation functions of renormalized operators must be UV finite.
   then we can find the renormalization factor by requiring
   finitness of the correlation function
   \eq{\label{crr}
   \vev{Z_\Phi^{1/2}\Phi_{j_1}(x_1)\ldots Z_\Phi^{1/2}\Phi_{j_L}(x_L)\,\,\OO^A_{\rm
   ren}(x)}.
   } 
    Here, $Z_\Phi$ is the wave-function renormalization factor, that is  
   multiplication 
   by $Z_\Phi$ makes the two-point correlator $\vev{\Phi_i\Phi_j}$
   finite. All renormalization factors depend on the UV cutoff $\La$ and 
on the 't~Hooft coupling in the large-$N$
   limit. By  standard arguments, the 
   renormalization factor determines
   the matrix of anomalous dimensions through
   \be\label{matad}
   \G=\frac{dZ}{d\ln\La}\cdot Z^{-1}.
   \ee
   Eigenvectors of $\G$ correspond to operators which are  multiplicatively
   renormalizable.
   The corresponding eigenvalues determine the anomalous dimensions
   of these operators.  Thus,
   \be
   \vev{\OO_n(x)\OO_n(y)}=\frac{\const}{|x-y|^{2(L+\gamma_n)}}
   \ee
   for the operator that corresponds to an eigenvector of $\G$ with an eigenvalue
   $\gamma_n$.

   How should one characterize the  Hilbert space\footnote{We shall
 call it a Hilbert space, even though it is finite-dimensional.} 
of  scalar operators of bare dimension $L$?
    Let us forget for
   a moment the
    cyclicity of the trace. Then in the natural basis
   \rf{natur} each operator is
   associated with an $SO(6)$ tensor with $L$ indices. 
   Such tensors form a $6^L$-dimensional linear space
   $\H=V_1\otimes\ldots\otimes V_L$, where $V_l=\mathbb{R}^6$ is associated with
   an $SO(6)$ index in the $l$th position in $ \psi^{i_1\ldots i_{l}\ldots
   i_{\d0}}$.
   The anomalous dimensions are thus eigenvalues of a $6^L\times 6^L$
   matrix.                  % which we want to diagonalize.
   It will prove extremely useful to regard $\H$ as a Hilbert space of a {\it spin
   system}.
   That is, let us consider a one-dimensional lattice with $L$ sites whose ends are
   identified and let each lattice site host a  six-dimensional 
   real vector.
   The space of states for such a spin system is isomorphic to $\H$. 
The matrix of 
   anomalous dimensions is a Hermitean operator in $\H$ and can be regarded 
   as a Hamiltonian of the spin system. Recalling that  wave functions
    %of the spin system 
   which
   differ by a cyclic permutation of indices correspond to the same
   operator,
   we should impose the
   constraint that physical states have zero total momentum:
   \eq{
   U\ket{\psi}=\ket{\psi},
   }
   where $U$ is the translation operator
   \eq{
   U\, a_1\otimes\ldots \otimes a_{L-1}\otimes a_L
   =a_L\otimes a_1\otimes\ldots \otimes a_{L-1}.
   }
   In the strict large-$N$ limit, all operators \rf{natur} are independent and there
   are
   no other constraints. 

   With the spin system interpretation in mind, let us compute the matrix of
   anomalous
   dimensions at one loop. The renormalization 
   of BMN operators with two impurities was extensively discussed, 
   so the essential pieces of the calculation for the 
   anomalous dimensions are present throughout the literature
(e.g. \cite{0202021,0205033,0205066,0205089}).
   We will therefore skip many
   details and give only salient features of the derivation,
   generalizing to arbitrary scalar operators. We 
   use the standard Feynman rules which follow from the
   Euclidean SYM action:
   \eq{
   S=\frac{1}{g^2}\int d^4x\,\tr\left\{
   \frac12\,F_{\mu\nu}^2+\br{D_\mu\Phi_i}^2-\frac12\,[\Phi_i,\Phi_j]^2
   +{\rm fermions}
   \right\},
   }
   and we will work in the Feynman gauge, in which the scalar and the gauge boson
   propagators are equal, up to Lorentz and  $SO(6)$ structures.

   \begin{figure}[h]
   %\hspace*{4cm}
   \begin{center}
   \epsfxsize=8cm
   \epsfbox{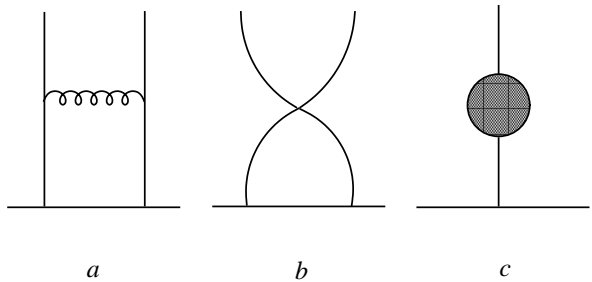}
   \end{center}
   \caption[x]{\small One-loop diagrams.}
   \label{diag}
   \end{figure}

   There are three types of planar 
   one-loop diagrams that contribute to the correlation function
   \rf{crr} (fig.~\ref{diag}). We depict the operator $\OO[\psi]$ by a horizontal
   bar
   with scalar propagators ending on each of the scalar fields  ({\it i.e.}
lattice
   sites) in the operator \rf{natur}. 
   Only lattice sites affected by loop corrections are shown in the figure.
   Since the gauge boson exchange is flavor-blind, the $Z$ factor associated with
    diagram (a) is diagonal in $SO(6)$ indices:
   $$
   Z^{(a)\ldots j_l j_{l+1}\ldots}_{\hphantom{(a)}\ldots i_l i_{l+1}\ldots}
   =I-\frac{\lambda}{16\pi^2}\,\ln\La\,\D^{j_l}_{i_l}\D^{j_{l+1}}_{i_{l+1}}.
   $$
   The $SO(6)$ structure of the $Z$ factor arising from  diagram (b) can be
   easily inferred from the structure of the quartic scalar vertex:
   $$
   \epsfxsize=4cm
   \epsfbox{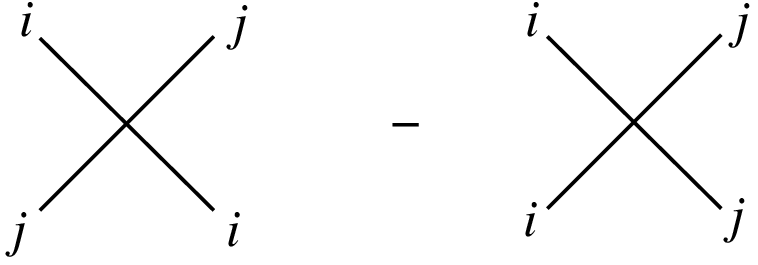}.
   $$
   Thus we find that
   $$
   Z^{(b)\ldots j_l j_{l+1}\ldots}_{\hphantom{(b)}\ldots i_l i_{l+1}\ldots}
   =I-\frac{\lambda}{16\pi^2}\,\ln\La\,
   \br{2\D^{j_{l+1}}_{i_l}\D^{j_l}_{i_{l+1}}-\D^{j_l}_{i_l}\D^{j_{l+1}}_{i_{l+1}}
   -\D_{i_l i_{l+1}}\D^{j_l j_{l+1}}}.
   $$
   
The one-loop self-energy correction in
    diagram (c) leads to the wave-function renormalization. 
   %of the scalar fields. 
   The corresponding renormalization
    factor was computed in  Feynman gauge 
   \cite{esz} and is given by
   $$
   Z_\Phi=1+\frac{\lambda}{4\pi^2}\,\ln\La.
   $$
   One half of the self-energy corrections in the correlation function \rf{crr}
   are cancelled by wave-function renormalization of the external legs. 
   The  remaining divergence
   should be cancelled by renormalization of the operator. The corresponding 
   $Z$ factor is proportional to the unit matrix, and can be written as
   $$
   Z^{(c)\ldots j_l j_{l+1}\ldots}_{\hphantom{(c)}\ldots i_l i_{l+1}\ldots}
   =I+\frac{\lambda}{8\pi^2}\,\ln\La\,\D^{j_l}_{i_l}\D^{j_{l+1}}_{i_{l+1}}.
   $$
   Adding all the pieces together, we find that the contribution from
each link of the
   lattice is
   \be\label{Zfactor}
   Z^{\ldots j_l j_{l+1}\ldots}_{\ldots i_l i_{l+1}\ldots}
   =I+\frac{\lambda}{16\pi^2}\,\ln\La\,
   \br{\D_{i_l i_{l+1}}\D^{j_l j_{l+1}}+2\D^{j_l}_{i_l}\D^{j_{l+1}}_{i_{l+1}}
   -2\D^{j_{l+1}}_{i_l}\D^{j_l}_{i_{l+1}}
   }.
   \ee
   The total $Z$ factor is the product over all links of the expression
in \refb{Zfactor}.

   The matrix of anomalous dimensions can be expressed in terms of two
   elemtary operators which act on each link: the trace
   operator, 
\be\label{trace}
K^{j_l j_{l+1}}_{i_l i_{l+1}}=\D_{i_l i_{l+1}}\D^{j_l j_{l+1}},
\ee
   and the permutation operator: 
   \be\label{perm}
P^{j_l j_{l+1}}_{i_l i_{l+1}}=\D^{j_{l+1}}_{i_l}\D^{j_l}_{i_{l+1}}.
\ee
 These
   operators act in the tensor product $\mathbb{R}^6\otimes\mathbb{R}^6$ 
   as
   \ar{
   K\,a\otimes b&=&(a\cdot b)\,\sum_i {\rm\widehat e}^i\otimes{\rm\widehat e}^i,
   \non
   P\,a\otimes b&=&b\otimes a,
   }
where ${\rm\widehat e}^i$ are a set of orthogonal unit vectors in 
$\mathbb{R}^6$.
   The matrix of anomalous dimensions is
   \eq{\label{spin}
   \G=\frac{\lambda}{16\pi^2}\sum_{l=1}^{L}
   \br{K_{l,l+1}+2-2P_{l,l+1}},
   }
   where the subscripts indicate that the operators act in the tensor
   product of  nearest-neighbor spins $V_l\otimes V_{l+1}$.
   By introducing the spin operators
   \eq{
   M_{ij}^{ab}=\D^a_i\D^b_j-\D^a_j\D^b_i
   }
   for each lattice site, we can rewrite the Hamiltonian in the form in which
    spin-spin interactions are manifest:
   \eq{
   \G=\frac{\lambda}{16\pi^2}\sum_{l=1}^{L}
   \left[
   M^{ab}_l M^{ab}_{l+1}-\frac{1}{16}\,\br{M^{ab}_l M^{ab}_{l+1}}^2+\frac{9}{4}
   \right].
   }

   The result in \rf{spin} for the matrix of anomalous dimensions in the form
   of a Hamiltonian of a spin system is the main result of this section.

   \newsection{Examples}

 The Hamiltonian in \refb{spin} posesses some remarkable properties. We
will see in the next section that it
   belongs to a unique series of integrable
   spin chains with $SO(n)$ symmetry. 
   For an arbitrary $SO(n)$ spin chain,
   integrability requires that the ratio of
   coefficients between the permutation operator
and the trace operator is $-(n/2-1)$. For $SO(6)$, this ratio is $-2$,
  precisely matching the ratio in \refb{spin}! 
Integrability allows
   one to use powerful techniques of the 
Bethe ansatz to diagonalize the Hamiltonian
   and  compute its eigenvalues. The review of the Bethe ansatz for the $SO(6)$
spin chain is given
in the next section.  

Since the Bethe ansatz utilizes
   rather sophisticated algebraic constructions, we would first
like to demonstrate
     the formalism by rederiving known results for some of the
simpler operators
   before invoking the Bethe ansatz machinery.

   The simplest and  most important scalar operators
   in \NN=4 SYM are chiral primaries,  
   operators which are
   symmetric and traceless in all $SO(6)$ indices. Chiral primaries
   are annihilated by the trace operator $K$ in \refb{trace}
   and are eigenstates of the permutation operator with an eigenvalue one.
   Therefore,
   \eq{
   \G\ket{ CPO}=0,
   } 
   which reflects the fact that
   scaling dimensions of chiral primaries are protected by
   supersymmetry and should not receive quantum corrections.

   Another interesting operator is the Konishi scalar,
   \eq{
   KO=\tr\Phi_i\Phi_i.
   }
   It is also invariant under permutations, but now the trace
   operator acts non-trivially: $K\ket{KO}=6\ket{KO}$. The Konishi operator
   corresponds to the lattice with two sites. Each link between the lattice 
   sites gives
   an equal contribution to the anomalous dimension, so 
   \eq{
   \G\ket{KO}=\frac{3\lambda}{4\pi^2}\,\ket{KO},
   }
   in agreement with the  calculation of \cite{9809192}.

   Consider now BMN operators with two impurities:
   \eq{
   \OO_{ij}=\sum_{l=0}^{J}\psi_l\,\tr\Phi_i Z^l \Phi_j Z^{J-l}~~~~~(i\neq
   j,~~i,j=3,\ldots,6).
   }
   The spin-chain Hamiltonian acts on such operators as
   a lattice Schr\"odinger operator with $\D'$ potential:
   \eq{
   \br{\G\psi}_l=-\frac{\lambda}{4\pi^2}
   \left[\psi_{l+1}+\psi_{l-1}-2\psi_l+\frac{1}{2}\br{\D_{l0}-\D_{lJ}}
   \br{\psi_0-\psi_J}\right].
   }
    The exact (multiplicatively renormalizable at any $J$) BMN operators
   with two impurities  were recently found by
   Biesert \cite{0211032}. His operators correspond to taking
   \eq{
   \psi^S_l=\cos\left[\frac{(2l+1)n\pi}{J+1}\right]
   }
   for states which are symmetric under interchange of $i$ and $j$,
   and 
   \eq{
   \psi^A_l=\sin\left[\frac{2(l+1)\pi n}{J+2}\right]
   }
    for antisymmetric states.
   It is straightforward to check that the above states
    are eigenfunctions of $\G$
   with eigenvalues
   \be\label{gSn}
   \gamma^S_n=\frac{\lambda}{\pi^2}\,\sin^2\br{\frac{\pi n}{J+1}}
   \ee
   and
   \be\label{gAn}
   \gamma^A_n=\frac{\lambda}{\pi^2}\,\sin^2\br{\frac{\pi n}{J+2}}.
   \ee
   As explained in \cite{0211032}, 
symmetric and antisymmetric operators with the same
   $J$ belong to different supermultiplets and for that reason their
   anomalous dimensions are different. 

   Finally, there are
   singlet BMN operators of the form:
   \eq{
   \OO=\sum_{l=0}^{J}\phi_l\,\sum_{i=3}^{6}\tr\Phi_i Z^l \Phi_i
   Z^{J-l}-\chi\tr\bar{Z}Z^{J+1}.
   } 
   The matrix of anomalous dimensions acts on these operators as
   \ar{
   \br{\G\phi}_l&=&-\frac{\lambda}{4\pi^2}
   \left[\phi_{l+1}+\phi_{l-1}-2\phi_l-\frac{1}{2}\br{\D_{l0}+\D_{lJ}}
   \br{\phi_0+\phi_J-\chi}\right].
   \non
   \G\chi&=&-\frac{\lambda}{4\pi^2}(\phi_0+\phi_J-\chi).
   }
   This is a Schr\"odinger operator with a self-consistent source and  a
   repulsive $\D$-function
    potential. Note that the source and the potential come from the trace
   term in the spin-chain Hamiltonian.
   Operators constructed in \cite{0211032} correspond to the wave functions
   \ar{
   \phi_l&=&\cos\left[\frac{(2l+3)\pi n}{J+3}\right],
   \non
   \chi&=&2\cos\br{\frac{\pi n}{J+3}}.
   }
   It is easy to check that they are eigenfunctions of
   the Hamiltonian with eigenvalues
   \be\label{gn}
   \gamma_n=\frac{\lambda}{\pi^2}\,\sin^2\br{\frac{\pi n}{J+3}},
   \ee
   in agreement with the anomalous dimensions computed in \cite{0211032}.

\sectiono{A short review of the Bethe ansatz equations}

In this section we review the Yang-Baxter equation, the construction
of commuting operators and the Bethe-ansatz  for an
$SO(n)$ chain where all sites in the chain transform in the vector
representation\footnote{For a nice explanation of the Yang-Baxter equation
and the algebraic Bethe ansatz see \cite{9605187}.}.

In order to find an integrable system, one
needs to construct an $R$-matrix.  An $R$-matrix
$R_{12}(u)$ acts on a tensor product of two $n$ dimensional vector spaces,
$V_1\otimes V_2$.  The parameter $u$ is the {\it spectral parameter} and
the matrix elements are explicitly given by  ${R_{12}(u)}^{i_1i_2}_{j_1j_2}$.
The {\it transfer matrix} $T(u)$ is constructed from the  $R$-matrix as
\be\label{transfer}
T(u)=R_{01}(u)R_{02}(u)R_{03}(u)...R_{0L}(u).
\ee
Here, the transfer matrix acts on the tensor product of $L+1$ 
$n$-dimensional vector spaces.  The sites on the chain are numbered from 
1 to $L$ while the space $V_0$ is an auxilary space.   One can think of
$T(u)$ as a matrix of operators that act on the $L$ sites of the chain, with
the different 
matrix elements given by ${T^{i_0}}_{j_0}(u)$.

If a system is integrable, then the $R$-matrix satisfies the
{\it Yang-Baxter equation}
\be\label{YBE}
R_{12}(u)R_{13}(u+v)R_{23}(v)=R_{23}(v)R_{13}(u+v)R_{12}(u)
\ee
where the three $R$-matrices act on the tensor product of three
$n$-dimensional vector spaces.
Given the Yang-Baxter equation, one can find the corresponding relation
for a product of transfer matrices
\be\label{YBEtr}
R_{ab}(u-v)T_a(u)T_b(v)=T_b(v)T_a(u)R_{ab}(u-v)
\ee
where the indices $a$ and $b$ refer to two different auxillary spaces,
but the transfer matrices act on the same chain of $L$ sites.  Writing
the components of the auxillary spaces explicitly, \refb{YBEtr} becomes
\be
{T_a^{i_a}}_{j_a}(u){T_b^{i_b}}_{j_b}(v)={{R^{-1}_{ab}}^{i_ai_b}}_{k_ak_b}(u-v)
{T_b^{k_b}}_{l_b}(v){T_a^{k_a}}_{l_a}(u){R_{ab}^{l_al_b}}_{j_aj_b}(u-v).
\ee
Taking the trace on the $V_a\otimes V_b$ tensor space, we get
\be
\Tr_a(T_a(u))\Tr_b(T_b(v))=\Tr_b(T_b(v))\Tr_a(T_a(u)),
\ee
and since the auxillary spaces are traced over, we can drop the labels $a$
and $b$ and write
\be\label{commrel}
[\Tr(T(u)),\Tr(T(v))]=0
\ee
for all $u$ and $v$.  For the case that we will be considering
it will turn out that these traces are order
$2L$ polynomials in
the spectral parameter, hence the Yang-Baxter equation implies that
there are up to $2L$ independent operators that are mutually commuting. 

Consider then the $R$-matrix acting on $V_1\otimes V_2$
\be\label{Rson}
R_{12}=\frac{1}{n-2}[u(2u+2-n)I_{12}-(2u+2-n)P_{12}+2uK_{12}],
\ee
where $I_{12}$, $P_{12}$ and $K_{12}$ are the identity, exchange
and trace operators  defined in the 
previous section.  This $R$-matrix satisfies the 
Yang-Baxter equation\cite{Resh1,Resh2}.  The verification
for this is straightforward, but tedious.  

Clearly, the 
transfer matrices will be polynomials of order $2L$ in $u$, which
we write as
\be\label{Texp}
T(u)=\sum_m u^m T_m.
\ee
The traces we write as
\be
t(u)\equiv\Tr(T(u))=\sum_m u^m t_m
\ee
 Let us find
the first few terms in the expansion.  Since $R_{12}(0)=P_{12}$, the
lowest order term in the expansion is
\be\label{T0}
T_0=\prod_{\ell=1}^LP_{0\ell}.
\ee
Hence the action of
this operator on the tensor product of the $L+1$ vector spaces is
\be
T_0 V_0\otimes V_1\otimes ...\otimes V_{L-1}\otimes V_{L}=
V_1\otimes V_2\otimes ...\otimes V_{L}\otimes V_0.
\ee
If we now take the trace over the $V_0$ space, we have that
\be
t_0V_1\otimes ...\otimes V_{L-1}\otimes V_{L}=
V_2\otimes ...\otimes V_{L}\otimes V_1.
\ee
Hence $t_0$ is the discrete shift operator, the operator that shifts
everything by one site. We already encountered this operator in
imposing the cyclicity of the trace on the SYM operators.

The next term in \refb{Texp} is found by replacing one $P_{0\ell}$ operator 
in \refb{T0} with 
\be
-I_{0\ell}-\frac{2}{n-2}P_{0\ell}+\frac{2}{n-2}K_{0\ell}
\ee
and summing over all positions $\ell$.   The contribution from
$P_{0\ell}$ will just give us the shift operator again.
To find the contributions of the other operators, note that
\be 
\Tr_0\left(\prod_{k=1}^{\ell-1}P_{0k}\prod_{k=\ell+1}^LP_{0k}\right)=t_0P_{\ell,\ell+1}\ee
and
\be
\Tr_0\left(\prod_{k=1}^{\ell-1}P_{0k}K_{0\ell}\prod_{k=\ell+1}^LP_{0k}\right)=
t_0K_{\ell,\ell+1}.
\ee
Hence we find that $t_1$ is given by
\be\label{t1eq}
t_1=\frac{2}{n-2}t_0\left(\sum_{\ell=1}^L(K_{\ell,\ell+1}-1-\frac{n-2}{2}P_{\ell,\ell+1})\right).
\ee
Since $t_1$ and $t_0$ are among a set of commuting operators and since
we are free to add a constant, we see
that
\be\label{t1t0}
\sum_{\ell=1}^L(K_{\ell,\ell+1}+\frac{n-2}{2}-\frac{n-2}{2}P_{\ell,\ell+1})
\ee
also commutes with these operators.

If we now consider the particular case of $SO(6)$, we see that \refb{t1t0}
is proportional to the anomalous dimension operator 
 in \refb{matad}!  Therefore, the one-loop
anomalous dimension operator described in the previous section can be mapped
to a Hamiltonian of an integrable system. 

Showing that a Hamiltonian is part of an integrable system is only part
of the story.  We also want to find the eigenstates and the  eigenvalues
of $t(u)=\Tr T(u)$.  In the Heisenberg spin chain, this is done most efficiently
by using the algebraic Bethe ansatz.
One can use the algebraic Bethe ansatz for the $SO(n)$ chain
as well \cite{dVK,9703023}.    However, as was
shown by Reshetikhin \cite{Resh1,Resh2}, there is another way
to find the eigenvalues of $t(u)$ which are
constrained by a series of Bethe equations.

Let us give a brief sketch of Reshetikhin's argument.
The first thing to observe is that the $R$-matrix in \refb{Rson} has
a crossing symmetry
\be\label{crossing}
(R_{12}(u))^{T_2}=R_{12}(-u+\frac{n-2}{2}),
\ee
where $T_2$ signifies a transpose on $V_2$ only.  Assuming that $u$ is
real, it is then straightforward to show that
\be
(t(u))^\dagger=t(-u+\frac{n-2}{2}).
\ee
Hence, the eigenvalues $\Lambda(u)$ of $t(u)$  satisfy
\be\label{crosseigen}
\bL(u)=\Lambda(-u+\frac{n-2}{2}).
\ee

Next consider the combination of $R$-matrices
\be
R_{12}(\frac{n-2}{2})R_{13}(u+\frac{n-2}{2})R_{23}(u)=K_{12}R_{13}(u+\frac{n-2}{2})R_{23}(u).
\ee
If we define $K_{12}^{\perp}$ as the orthogonal complement to the 
trace operator, then by the Yang-Baxter equation we have that
\be\label{fact1}
K_{12}R_{13}(u+(n-2)/2)R_{23}(u)K_{12}^{\perp}=0.
\ee
This means that $R_{13}(u+\frac{n-2}{2})R_{23}(u)$ can be written in lower
triangular form on the $V_1\otimes V_2$ space, where the upper left
block corresponds to the operator $K_{12}R_{13}(u+\frac{n-2}{2})R_{23}(u)K_{12}$
and the right lower block to $K_{12}^{\perp}R_{13}(u+\frac{n-2}{2})R_{23}(u)K_{12}^{\perp}$

We next note that 
\ben\label{fact2}
&&{R_{13}(u+\frac{n-2}{2})^{i_1i}}_{j_1k}{R_{23}(u)^{i_1k}}_{j_1j}
=\nonumber\\
&&\qquad\qquad\frac{1}{(n-2)^2}\Big[(4u^2-(n-2)^2)({A^{i_1i}}_{j_1j}
+u^2{B^{i_1i}}_{j_1j})+4u^2{C^{i_1i}}_{j_1j}\Big]
\een
where only the $k$ index is summed over and where
\begin{eqnarray}\label{fact3}
{A^{i_1i}}_{j_1j}&=&-\delta^{i_1i}{\delta^{i_1}}_{j}\nonumber\\
{B^{i_1i}}_{j_1j}&=&{\delta^{i_1}}_{j_1}{\delta^{i}}_{j}
\nonumber\\
{C^{i_1i}}_{j_1j}&=&n{\delta^{i_1}}_{j_1}\delta^{i_1i}\delta_{j_1j}
-{\delta^{i}}_{j_1}\delta_{j_1j}.
\end{eqnarray}
One can then show by using the independence of ${A^{i_1i}}_{j_1j}$
on $j_1$ that
\be\label{fact4}
\sum_{j_1}{A^{i_1i}}_{j_1j}{C^{j_1I}}_{k_1 J}=0.
\ee
Finally, we note that
\be\label{fact5}
{R_{13}(\frac{n-2}{2})^{i_1i}}_{j_1k}{R_{23}(0)^{i_2k}}_{j_2j}
=0\qquad{\rm if}\qquad  {i_1\ne i_2\atop j_1\ne j_2}.
\ee

Putting together the relations in \refb{fact1}--\refb{fact5} and using
the relation in \refb{crosseigen}, one can then
show that
\be\label{u-urel}
\Lambda(u)\bL(-u)=\frac{1}{(n-2)^{2L}}(u^2-1)^L(4u^2-(n-2)^2)^L+u^L\Lambda_r(u)
\ee
where $\Lambda_r(u)$ is a remainder term that is yet to be determined.
The relation in \refb{u-urel} is highly constraining.  As was shown 
by Reshetikhin \cite{Resh1,Resh2}, its solution is
\begin{eqnarray}\label{Lsol}
\Lambda(u)=\frac{1}{(n-2)^L}\Big[(u-1)^L(2u-n+2)^LH(u)&+&u^L(2u-n+4)^LF(u)
\nonumber\\
&+&u^L(2u-n+2)^LG(u)\Big]
\end{eqnarray}
where in order to satisfy \refb{u-urel} and crossing symmetry
\begin{eqnarray}
H(u)\overline H(-u)&=&1\nonumber\\
\overline F(-u+\frac{n-2}{2})&=&H(u)\nonumber\\
\overline G(-u+\frac{n-2}{2})&=&G(u).
\end{eqnarray}
A solution for the first of these equations is
\be\label{Hsol}
H(u)=\prod_{j=1}^{n_1}\frac{u-iu_{1,j}+1/2}{u-iu_{1,j}-1/2}
\ee
where the number $n_1$ and the possible values $u_{1,m}$ will depend on the particular
eigenstate.  If  $u_{1,m}$ is complex, then its conjugate must also
be contained in the product.  

The function $G(u)$ will be written as a sum
\be
G(u)=\sum_{q=1}^{n-2}G_q(u)
\ee
where 
\be
\overline G_{n-1-q}(-u+\frac{n-2}{2})=G_q(u)
\ee
Let us assume that $n=2k$.  Then the various $G_q(u)$ are given by
\begin{eqnarray}
G_q(u)&=&\prod_{j=1}^{n_q}\frac{u-iu_{q,j}-q/2-1}{u-iu_{q,j}-q/2}
\prod_{j=1}^{n_{q+1}}\frac{u-iu_{q+1,j}-q/2+1/2}{u-iu_{q+1,j}-q/2-1/2}\qquad
1\le q<k-2
\nonumber\\
G_{k-2}(u)&=&\prod_{j=1}^{n_{k-2}}\frac{u-iu_{k-2,j}-k/2}{u-iu_{q,j}-k/2+1}
\prod_{j=1}^{n_{k-1}}\frac{u-iu_{k-1,j}-k/2+3/2}{u-iu_{k-1,j}-k/2+1/2}
\prod_{j=1}^{n_{k}}\frac{u-iu_{k,j}-k/2+3/2}{u-iu_{k,j}-k/2+1/2}
\nonumber\\
G_{k-1}(u)&=&
\prod_{j=1}^{n_{k-1}}\frac{u-iu_{k-1,j}-k/2+3/2}{u-iu_{k-1,j}-k/2+1/2}
\prod_{j=1}^{n_{k}}\frac{u-iu_{k,j}-k/2+1/2}{u-iu_{k,j}-k/2+1/2}
\end{eqnarray}

However, the eigenvalues must be a polynomial in $u$, but given the
structure of the above functions, it appears that $\Lambda(u)$ will have poles
at $u=iu_{q,m}$ for all of the various values of $q$ and $m$.  Hence,
there has to be intricate relations between the different values
$u_{q,m}$ in order that the poles cancel.  The relations were derived in
\cite{Resh1,Resh2} and are given by
\begin{eqnarray}\label{betheeqs}
&&\left(\frac{u_{1,i}+i/2}{u_{1,i}-i/2}\right)^L=
\prod_{j\ne i}^{n_1}\frac{u_{1,i}-u_{1,j}+i}{u_{1,i}-u_{1,j}-i}
\prod_{j}^{n_2}\frac{u_{1,i}-u_{2,j}-i/2}{u_{1,i}-u_{2,j}+i/2}
\nonumber\\
&&1=\prod_{j\ne i}^{n_q}\frac{u_{q,i}-u_{q,j}+i}{u_{q,i}-u_{q,j}-i}
\prod_{j}^{n_{q-1}}\frac{u_{q,i}-u_{q-1,j}-i/2}{u_{q,i}-u_{q-1,j}+i/2}
\prod_{j}^{n_{q+1}}\frac{u_{q,i}-u_{q+1,j}-i/2}{u_{q,i}-u_{q-1,j}+i/2}
\qquad 1<q<k-2\nonumber\\
&&1=\prod_{j\ne i}^{n_{k-2}}\frac{u_{k-2,i}-u_{k-2,j}+i}{u_{k-2,i}-u_{k-2,j}-i}
\prod_{j}^{n_{k-3}}\frac{u_{k-2,i}-u_{k-3,j}-i/2}{u_{k-2,i}-u_{k-3,j}+i/2}
\nonumber\\
&&\qquad\qquad\qquad\qquad\times
\prod_{j}^{n_{k-1}}\frac{u_{k-2,i}-u_{k-1,j}-i/2}{u_{k-2,i}-u_{k-1,j}+i/2}
\prod_{j}^{n_{k}}\frac{u_{k-2,i}-u_{k,j}-i/2}{u_{k-2,i}-u_{k,j}+i/2}\nonumber\\
&&1=\prod_{j\ne i}^{n_{k-1}}\frac{u_{k-1,i}-u_{k-1,j}+i}{u_{k-1,i}-u_{k-1,j}-i}
\prod_{j}^{n_{k-2}}\frac{u_{k-1,i}-u_{k-2,j}-i/2}{u_{k-1,i}-u_{k-2,j}+i/2}
\nonumber\\
&&1=\prod_{j\ne i}^{n_{k}}\frac{u_{k,i}-u_{k,j}+i}{u_{k,i}-u_{k,j}-i}
\prod_{j}^{n_{k-2}}\frac{u_{k,i}-u_{k-2,j}-i/2}{u_{k,i}-u_{k-2,j}+i/2}
\end{eqnarray}
These are the analogs of the Bethe equations for the Heisenberg spin chain
\cite{Bethe},
and the solutions are often called the Bethe roots.
It was subsequently shown, that these series of equations can be
generalized to arbitrary groups in different representations \cite{ow}.  
The generalized equations are given by
\be\label{bethegen}
\left(\frac{u_{q,i}+i\vec\al_q\cdot \vec w/2}{u_{q,i}-i\vec\al_q\cdot \vec w/2}\right)^L=
\prod_{j\ne i}^{n_q}\frac{u_{q,i}-u_{q,j}+i\vec\al_q\cdot\vec\al_q/2}{u_{q,i}-u_{q,j}-i\vec\al_q\cdot\vec\al_q/2}
\prod_{q'\ne q}\prod_{j}^{n_{q'}}\frac{u_{q,i}-u_{q',j}+i\vec\al_q\cdot\vec\al_{q'}/2}{u_{q,i}-u_{q',j}+i\vec\al_q\cdot\vec\al_{q'}/2}.
\ee
 The
different parameters $u_{q,i}$ are associated with the simple roots
of the Lie group $\vec\al_q$, and  the factor on the left hand side of
the equations depend on the maximum weight of the representation, $\vec w$.
In the case of $SO(2k)$ in the vector representation, we see that 
\refb{betheeqs} has the form in \refb{bethegen}.  Finally, for the particular
case of $SO(6)$ where $k-2=1$, 
the first equation in \refb{betheeqs} is modified to\footnote{The simple roots 
of $SO(6)$ are $\vec\alpha_1=(1,-1,0)$, $\vec\alpha_2=(0,1,-1)$, $\vec\alpha_3=(0,1,1)$, and
the weight of the vector representation is $\vec{w}=(1,0,0)$.}
\be\label{betheSO6}
\left(\frac{u_{1,i}+i/2}{u_{1,i}-i/2}\right)^L=
\prod_{j\ne i}^{n_1}\frac{u_{1,i}-u_{1,j}+i}{u_{1,i}-u_{1,j}-i}
\prod_{j}^{n_2}\frac{u_{1,i}-u_{2,j}-i/2}{u_{1,i}-u_{2,j}+i/2} 
\prod_{j}^{n_3}\frac{u_{1,i}-u_{3,j}-i/2}{u_{1,i}-u_{3,j}+i/2}.
\ee 
The other two equations read
\ar{\label{betheso}
&&1=\prod_{j\ne i}^{n_{2}}\frac{u_{2,i}-u_{2,j}+i}{u_{2,i}-u_{2,j}-i}
\prod_{j}^{n_{1}}\frac{u_{2,i}-u_{1,j}-i/2}{u_{2,i}-u_{1,j}+i/2}
\nonumber\\
&&1=\prod_{j\ne i}^{n_{3}}\frac{u_{3,i}-u_{3,j}+i}{u_{3,i}-u_{3,j}-i}
\prod_{j}^{n_{1}}\frac{u_{3,i}-u_{1,j}-i/2}{u_{3,i}-u_{1,j}+i/2}\,.
}

Now from \refb{Lsol} and \refb{Hsol} we can find the eigenvalues of the shift
operator and the Hamiltonian.  The eigenvalues of the shift operator are
\be
\Lambda(0)=H(0)=\prod_{i=1}^{n_1}\frac{u_{1,i}+i/2}{u_{1,i}-i/2}.
\ee
Hence the momenta of the eigenstates is
\be\label{mom}
P=-i\log(\Lambda(0))=-i\sum_i^{n_1}\log\frac{u_{1,i}+i/2}{u_{1,i}-i/2}
=\sum_i^{n_1}p(u_{1,i}).
\ee
The corresponding energies are
found from the eigenvalues of $t_1$, $\Lambda_1$, which are
\be
\Lambda_1=\frac{d}{du}H(u)\Big|_{u=0}-L\frac{n}{n-2}H(0).
\ee
Using \refb{t1eq} and \refb{t1t0}, we see that the energy eigenvalues
are
\be\label{energy}
E=\frac{n-2}{2H(0)}\frac{d}{du}H(u)\Big|_{u=0}=\frac{n-2}{2}
\sum_i^{n_1}\eps(u_{1,i}).
\ee
where
\be\label{perel}
\eps(u)=-\frac{d}{du}p(u)=4\sin^2\left(\frac{p(u)}{2}\right)
=\frac{1}{u^2+1/4}\,.
\ee
Hence the parameters $u_{1,i}$ are rapidity parameters for particle like
excitations of the ground state.  

Thus, specializing to $SO(6)$ and
using \refb{matad}, \refb{t1t0} and \refb{energy}, 
we find that the corresponding anomalous
dimension is
\be\label{anomdim}
\gamma=\frac{\lambda}{8\pi^2}
\sum_{i=1}^{n_1}\eps(u_{1,i}).
\ee

\sectiono{Applying the Bethe ansatz}

In this section we apply the results of the previous sections to many 
different scenarios.  Sometimes we will reduce our space of operators
to those involving just $Z$ and $W$ scalar fields\footnote{It is useful to
combine six real fields into three complex scalars:
$Z=\Phi_1+i\Phi_2$, $W=\Phi_3+i\Phi_4$, $Y=\Phi_5+i\Phi_6$, which
can be regarded as lowest components of three chiral superfields
in the ${\cal N}=1$ formalism.}.  In this case our
problem is basically reduced to a Heisenberg spin chain. Including
other fields complicates the problem somewhat, but we are still able
to make many statements about the eigenvalues.

As we saw in the previous section, the $SO(6)$ chain has three types
of excitations, with each type associated with one of the simple roots
of the $SO(6)$ Dynkin diagram.  Those associated with $\al_1$ are on
a somewhat different footing than those associated with $\al_2$ and
$\al_3$, since only the $\al_1$ excitations carry momentum and
energy.  However, the other two types of excitations can indirectly affect
the energy of the state by modifying the $u_{1}$ rapidities.

If we were to limit ourselves to only $u_{1}$ excitations, then we
see that the Bethe ansatz equations in \refb{betheSO6} reduce to that
of the ordinary Heisenberg spin chain.  For this case, the different
lattice sites can have one of two values (spin up or down).  The Heisenberg
spin chain has no trace term either, so the corresponding situation
for the operator chains is to have two types of fields where the trace
term does not contribute.  So for example, we could have chains made up
of $Z$ and $W$ terms only.  If we call the
ground state $\tr Z^J$, then the particle excitations with rapidities
$u_{1,i}$ create $W$ operators in the chain.  Another way to see this
is that the $Z$ field is the highest weight in the vector representation
of $SO(6)$, which we write as $\vec\mu_1=(1,0,0)$. 
Subtracting an $\vec\al_1=(1,-1,0)$ root
then gives $(0,1,0)$ which corresponds to the $W$ field.  

Now suppose that we were to try and create $u_{2}$ and $u_3$ excitations
without any $u_1$ excitations.  It is not too hard to see from the Bethe
equations that this is not possible.  This is clear from the perspective
of the group representations as well, since $\vec\mu_1-\vec\al_2$ and
$\vec\mu_1-\vec\al_3$ are not $SO(6)$ weights.  However
$\vec\mu_1-\vec\al_1-\vec\al_2$ and $\vec\mu_1-\vec\al_1-\vec\al_2$ are
weights, so given some $u_1$ excitations, it is possible to have
$u_2$ and $u_3$ excitations.

We should also note that our $SO(6)$ lattice chain 
appears in a trace, which means that
the corresponding wave functions are invariant under translation.  Hence
the total momentum is zero.   So in all considerations we require the
trace condition for the $u_{1,i}$
\be\label{tracec}
\prod_{i=1}^{n_1}\frac{u_{1,i}+i/2}{u_{1,i}-i/2}=1.
\ee

\subsection{Two impurities}

We first consider the case of two impurities, that is 
two $u_1$ excitations, which we label as $u_{1,1}$ and $u_{1,2}$.  
We need at least two impurities if we want to have excitations with non-zero
momentum, but with zero total momentum to satisfy the trace condition.
With two impurities the bare dimension exceeds the R charge
by two units: $L=J+2$.  From \refb{mom}
we have that
\be\label{2imps}
\frac{u_{1,1}+i/2}{u_{1,1}-i/2}\frac{u_{1,2}+i/2}{u_{1,2}-i/2}=1.
\ee
Recalling that $u_{1,2}=u^*_{1,1}$ unless they are both real, we see
that the only solutions have $u_{1,2}=-u_{1,1}$ with both values real.
Now using \refb{betheSO6}, we find
\be
\left(\frac{u_{1,1}+i/2}{u_{1,1}-i/2}\right)^L=\frac{2u_{1,1}+i}{2u_{1,1}-i}
\ee
and so we find that 
\be
p(u_{1,1})=\frac{2\pi n}{L-1}=\frac{2\pi n}{J+1}
\ee
and from \refb{perel}
\be
\eps(u_{1,1})=\frac{1}{u_{1,1}^2+1/4}=4\sin^2\frac{\pi n}{J+1}.
\ee
Therefore, using \refb{perel} and \refb{anomdim}
the anomalous dimension for this configuration is
\be
\gamma^S_n=\frac{\lambda}{16\pi^2}\frac{6-2}{2}\times 2\eps(u_{1,1})
=\frac{\lambda}{\pi^2}\sin^2\frac{\pi n}{J+1},
\ee
which agrees with the result in \refb{gSn}.  With no 
$u_2$ or $u_3$ excitations, 
the impurities are both $W$'s and so their representation is
symmetric traceless.

On top of the $u_1$ impurities, we can also add up to one each of
the $u_2$ and $u_3$ impurities in a nontrivial way.  
Putting in a $u_2$ impurity, we see
that  \refb{2imps} is unchanged, so $u_{1,1}=-u_{1,2}$.  Using \refb{betheso},
we also have that
\be\label{u2eq}
1=\frac{u_2-u_{1,1}-i/2}{u_2-u_{1,1}+i/2}\frac{u_2+u_{1,1}-i/2}{u_2+u_{1,1}+i/2}.
\ee
The only solutions to this are $u_2=\infty$ and $u_2=0$.  The first case
is the trivial solution in that it gives us the same anomalous dimension
as before.  This corresponds to having a $W$ and a $Y$ 
in the symmetric representation.  Taking the second solution and plugging
it into \refb{betheSO6}, we find that
\be
p(u_{1,1})=\frac{2\pi n}{L}=\frac{2\pi n}{J+2},
\ee
and the anomalous dimension is
\be\label{anomAn}
\gamma^A_n=\frac{\lambda}{\pi^2}\sin^2\frac{\pi n}{J+2},
\ee
the result previously given in \refb{gAn}.  This then is the antisymmetric
combination of $W$ and $Y$.  This is part of the  self-dual
representation of the $SO(4)$ subgroup.

If we now also add a $u_3$ impurity, then $u_3$ has an equation identical
to that for $u_2$ in \refb{u2eq}.  If there is no $u_2$ impurity, then
the anomalous dimension is the same as in \refb{anomAn}.  This is part of
 the anti-selfdual representation of $SO(4)$.
With both types of impurities, the nontrivial solutions then have
$u_2=u_3=0$ and so \refb{betheSO6} gives
\be
p(u_{1,1})=\frac{2\pi n}{L+1}=\frac{2\pi n}{J+3},
\ee
and the anomalous dimension is
\be
\gamma_n=\frac{\lambda}{\pi^2}\sin^2\frac{\pi n}{J+3},
\ee
the result previously given in \refb{gn}.  Notice that $-\vec\al_2$ takes
$W$ to $Y$ and $-\vec\al_3$ takes the $W$ to $\overline Y$.  But we also have
that $-\vec\al_2-\vec\al_3$ takes $W$ to $\overline W$ and that
$-2\vec\al_1-\vec\al_2-\vec\al_3$ takes a $Z$ to $\overline Z$.  Hence this
last result corresponds to the $SO(4)$ invariant of the two impurities.

\subsection{More than two impurities}

In this section we consider the addition of many impurities and compute
their anomalous dimensions, up to first order in $1/J$.  For the most part
we will limit our discussion to having only $u_1$ excitations.   Hence,
these will only be a subset of possible $SO(6)$ representations, namely,
the real representations with $2L$ boxes in
the $SU(4)$ Young Tableaux.
At the end of the section we will discuss the addition of a single $u_2$
or $u_3$ impurity.

Once we have more than two impurities, it is now possible to have complex
$u_1$ rapidities.  In fact, this possibility is basically forced on us
when we want to find BMN states where a particular oscillator appears more
than once. In the BMN limit, the momenta of the excitations should be
small, and so the phases in the Bethe equations are small.  But if two
excitations have identical momenta, then the combination 
$\frac{u_{1,1}- u_{1,2}+i}{u_{1,1}- u_{1,2}-i}$ which appears in the righthand side of the
Bethe equations will have a large phase. 

The resolution of this problem is that $u_{1,1}$ and $u_{1,2}$ get imaginary
pieces such that $u^*_{1,2}=u_{1,1}$.  This way we can get a small phase
so long as $|{\rm Im}u_{1,1}|\gg 1.$  The individual momenta of the excitations
are complex, but the combined momentum
\be
p(u_{1,1},u_{1,2})=p(u_{1,1})+p(u_{1,2})=-i\log\frac{u_{1,1}+i/2}{u_{1,1}-i/2}
-i\log\frac{u^*_{1,1}+i/2}{u^*_{1,1}-i/2}
\ee
 is real.  Note further that the combined energy from these two excitations
is
\be
\eps(u_{1,1})+\eps(u_{1,2})=4\sin^2\left(\frac{p(u_{1,1})}{2}\right)+
4\sin^2\left(\frac{p(u_{1,2})}{2}\right)\le8\sin^2\left(\frac{p(u_{1,1},u_{1,2})}{4}\right)
\ee
where there is an equality only if the individual momenta are real. Hence,
this configuration corresponds to a bound state of two particles\footnote{If
the momentum were of order 1, then the separation between $u_{1,1}$ and 
$u_{1,2}$ would be close to $i$.
In
the literature, these bound states are called ``strings'', but we will stick to
calling them bound states for obvious reasons.}, since the
combined energy is less than twice the energy of a single particle with
momentum $p(u_{1,1},u_{1,2})/2$.  This can be generalized to many particles
as well, where the individual momenta are complex, but their sum is real.
So a BMN state with $M$ oscillators at the same level would correspond to
a bound state of $M$ particles.

Unfortunately, it does not seem possible to find exact generic solutions to the
Bethe equations for more than two excitations.  However, it is possible to
at least find $1/J$ corrections in the BMN limit.
If we have particles with small momenta, then the values of $u_{1,i}$ are
large.  From the Bethe equations, we see  to leading order that these
are
\be
u_{1,n}\approx \frac{L}{2\pi k_n}
\ee
where $k_n$ is an integer.  Allowing for bound states, let us group the
various excitations as $\mu_i^{(n)}$, where 
\be\label{muansatz}
\mu_i^{(n)}=\frac{1}{2\pi k_n}(L+iL^{1/2}\nu_i^{(n)}+\delta_i^{(n)})+{\rm O}(L^{-1/2}).
\ee
We assume that $k_n\ne k_{m}$ if $n\ne m$ and the index $i$ sums over
the $M_n$ particles making up the bound state at $k_n$.
We can now expand the Bethe equations in  \refb{betheSO6}, \rf{betheso} 
in powers of
$1/\sqrt{L}$.  Solving for the  zeroth order term in the expansion gives integer
$k_n$.  

Next solving for the $L^{-1/2}$ term in the expansion gives the
equation
\be\label{nueq}
\nu_i^{(n)}=\sum_{j\ne i}\frac{2}{\nu_i^{(n)}-\nu_j^{(n)}}.
\ee
It turns out that we don't need to explicitly know the $\nu_i^{(n)}$ when
computing the anomalous dimension to this order, but
let us consider solutions of \refb{nueq} for a few different values of $M_n$
anyway.
If $M_n=2$, then we have that $\nu_1^{(n)}=-\nu_2^{(n)}=1$.  If
$M_n=3$, then $\nu_1^{(n)}=-\nu_2^{(n)}=\sqrt{3}$ and $\nu_3^{(n)}=0$.
Finally, let us consider the case where $M_n\gg 1$, then we can
describe the distribution of $\nu^{(n)}$'s by a continuous density and
may approximate
the sum by an integral
\be\label{wigner}
\nu=P\int_{-a}^{a}d\nu'\ \frac{2\ \rho^{(n)}(\nu')}{\nu-\nu'}
\ee
which is Wigner's equation for the eigenvalues of a large $N$ Hermitian
matrix model with a Gaussian potential.  Standard techniques give
\be
\rho^{(n)}(\nu)=\frac{1}{2\pi}\sqrt{a^2-\nu^2}\qquad\qquad a=2\sqrt{M_n}.
\ee
Notice that if $M_n\sim L$, then the maximum value of $\nu^{(n)}_i \sim \sqrt{L}$
and so the ansatz in \refb{muansatz} breaks down.

Next solving for the $L^{-1}$ term in the expansion of the Bethe equations 
leads to the equation
\be\label{deeq}
\delta_i^{(n)}+(\nu_i^{(n)})^2+
2\sum_{j\ne i}\frac{\delta_i^{(n)}-\delta_j^{(n)}}
 {(\nu_i^{(n)}-\nu_j^{(n)})^2}+2\sum_{m\ne n} M_m\frac{k_m}{k_m-k_n}=0.
\ee
To solve this equation, we make the ansatz that
\be
\delta_i^{(n)}=c_n (\nu_i^{(n)})^2+b_n.
\ee
Substituting this back into \refb{deeq} and making use of \refb{nueq}
 we find that
\begin{eqnarray}
c_n&=&-\frac{1}{3}\nonumber\\
b_n&=&=-\frac{2}{3}(M_n-1)-2\sum_{m\ne n}\frac{M_mk_m}{k_m-k_n},
\end{eqnarray}
and so
\be
\delta_i^{(n)}=-\frac{1}{3}(\nu_i^{(n)})^2
-\frac{2}{3}(M_n-1)-2\sum_{m\ne n}\frac{M_mk_m}{k_m-k_n}.
\ee

Let us now place these results into the energies in \refb{perel} and 
\refb{energy}.  Up to
and including corrections of order $1/L$, we can approximate these as
\be
\eps_i^{(n)}=\frac{1}{(u_i^{(n)})^2}.
\ee
Hence, the energy coming from a single bound state is
\begin{eqnarray}\label{epsneq}
\eps^{(n)}&=&\sum_{i=1}^{M_n}\eps_i^{(n)}
=\left(\frac{2\pi k_n}{L}\right)^2\sum_i\left[1-\frac{2}{L}\D_i^{(n)}
-\frac{3}{L}(\nu_i^{(n)})^2\right]\nonumber\\
&=&
\left(\frac{2\pi k_n}{L}\right)^2\frac{1}{L}\left[LM_n+\frac{4}{3}M_n(M_n-1)
+4\sum_m\frac{M_nM_mk_m}{k_m-k_n}-\frac{7}{3}\sum_i(\nu_i^{(n)})^2\right].
\end{eqnarray}
Using \refb{nueq}, we have that
\be
\sum_i(\nu_i^{(n)})^2=M_n(M_n-1).
\ee
Putting this back in \refb{epsneq} we find 
\be\label{epsneq2}
\eps^{(n)}=
\left(\frac{2\pi k_n}{L}\right)^2\frac{M_n}{L}\left[L-(M_n-1)
+4\sum_{m\ne n}\frac{M_mk_m}{k_m-k_n}\right].
\ee
The negative term inside \refb{epsneq2} is basically the contribution of
the binding energy, where the binding energy is present if $M_n>1$.
The last term in \refb{epsneq2} comes from interactions among
the different bound states.

The anomalous dimension is then found by adding up the $\eps^{(n)}$, giving
\be\label{anomgen}
\gamma=\frac{\lambda}{2L^3}\sum_n M_nk_n^2\left[L-(M_n-1)
+4\sum_{m\ne n}\frac{M_mk_m}{k_m-k_n}\right]+{\rm O}(L^{-4}).
\ee
The trace condition in \refb{tracec} requires that
\be
\sum_n M_nk_n=0.
\ee
We can then use this to reduce \refb{anomgen} to
\be\label{anomgen2}
\gamma=\frac{\lambda}{2L^3}\sum_n M_nk_n^2(L+M_n+1)+{\rm O}(L^{-4}).
\ee
Since we have added $n_1$ impurities, $L=J+n_1$.  Writing $\gamma$ in
terms of $J$ we find
\be
\gamma=\frac{\lambda}{2J^3}\sum_n M_nk_n^2(J-2n_1+M_n+1)
+{\rm O}(J^{-4}).
\ee

Let us now add a single  $u_2$ and/or a single $u_3$ impurity to the mix.
We have not yet found an explicit formula analagous to \refb{anomgen}
for a generic number of these impurities.  With only one each of
these impurities, the Bethe equations lead to 
\be
1=\prod_{i=1}^{n_1}\frac{u_2-u_{1,i}-i/2}{u_2-u_{1,i}+i/2},
\ee
and  an identical equation for $u_3$.  Hence, $u_2$ and $u_3$ can
be determined in terms of $u_{1,i}$ by solving an order $n_1$ polynomial
equation.  This can be solved if $n_1$ is small, but in general the
equation appears complicated.  However, it is easy to see using the
trace condition in \refb{tracec} that $u_2=0$ and $u_3=0$ are always
solutions to the equations.  

If we only have one of the impurities, with its value set to $0$, then
we see that the Bethe equation in \refb{betheSO6} is identical to the
equation with no impurities, except that $L$ is replaced with $L+1$.  So
if the exact solution could be found for the case with only $u_1$ impurities,
then the solution would be known for this case as well.  Likewise, if
we have both a $u_2=0$ and a $u_3=0$ impurity, then we should replace
$L$ by $L+2$.  

\sectiono{Large excitations}

Ultimately, we would like to solve string theory for the full $AdS_5\times
S_5$ and not just for the plane wave limit.  Then one could compare
the anomalous dimensions of all gauge invariant operators.  Likewise,
one would need to actually compute the dimensions of these operators
in the field theory.
So far, we have been restricting
ourselves to large $R$-charge, where we are limited to finding $1/J$ 
corrections to  BMN operators.  We can think of this as the dilute
gas limit \cite{0202021}.

Remarkably, the Bethe ansatz equations can be used to ascertain information about
operators outside the BMN regime of validity.  For example, one might ask
what is the largest possible anomalous dimension for an operator (made up
of scalars only)
with engineering dimension $L$.
This should be an $SO(6)$ singlet.  It turns out that this is solvable
in the large $L$ limit \cite{Resh3},
with a solution similar to that
of the ground state of the Heisenberg anti-ferromagnet.  Of course
the ground state of the Heisenberg anti-ferromagnet also corresponds
to a particular $SO(6)$ representation.

We now review the solution in \cite{Resh3} and then use the result to find
the anomalous dimension.
To find the solution, 
we assume a large number of excitations of all
impurity types.  To maximize the energy, we should take the maximal
number of impurities such that the solutions to the Bethe equations are
all real.  
If we take the log of \refb{betheSO6} and 
 \refb{betheso}, we find the equations
\begin{eqnarray}\label{ataneqs}
L\vartheta(2u_{1,j})&=&j\pi+\sum_{j\ne i}\vartheta(u_{1,i}-u_{1,j})
-\sum_j\vartheta(2(u_{1,i}-u_{2,j}))-\sum_j\vartheta(2(u_{1,i}-u_{3,j}))
\nonumber\\
0&=&j\pi+\sum_{j\ne i}\vartheta(u_{2,i}-u_{2,j})
-\sum_j\vartheta(2(u_{2,i}-u_{1,j}))
\nonumber\\
0&=&j\pi+\sum_{j\ne i}\vartheta(u_{3,i}-u_{3,j})
-\sum_j\vartheta(2(u_{3,i}-u_{1,j}))
\end{eqnarray}
where
\be
\vartheta(u)=\arctan(u).
\ee
Since the wave functions would be zero if $u_{1,i}=u_{1,j}$ with $i\ne j$,
the various roots are pushed onto different branches of the arctangent.

If $L$ is very large, then we can replace $j/L$ by a continuous variable $x$
and the Bethe roots by $u_1(x)$, $u_2(x)$ and $u_3(x)$.  
By symmetry, we expect  the distribution of $u_{2,i}$ and $u_{3,i}$ to
be identical, hence we set $u_2(x)=u_3(x)$.  The equations in
 \refb{ataneqs} now become
\begin{eqnarray}
\vartheta(2u_1(x))&=&\pi x +\int dy\vartheta(u_1(x)-u_1(y))-
2\int dy\vartheta(2(u_1(x)-u_2(y))\nonumber\\
0&=&\pi x +\int dy\vartheta(u_2(x)-u_2(y))-
\int dy\vartheta(2(u_2(x)-u_1(y)).
\end{eqnarray}
We now take derivatives with
respect to $u_1(x)$ and $u_2(x)$, which gives us
\ben\label{contbe}
\frac{2}{4u^2+1}&=&\pi\rho_1(u)+\int_{-\infty}^\infty du'
\frac{\rho_1(u')}{(u-u')^2+1}-4\int_{-\infty}^\infty du'
\frac{\rho_2(u')}{4(u-u')^2+1}\nonumber\\
0&=&\pi\rho_2(u)+\int_{-\infty}^\infty du'
\frac{\rho_2(u')}{(u-u')^2+1}-2\int_{-\infty}^\infty du'
\frac{\rho_1(u')}{4(u-u')^2+1},
\een
where the $\rho_1(u)$ and $\rho_2(u)$ are the root densities
\be
\rho_1(u)=\frac{dx}{du_1(x)}\Big|_{u_1(x)=u}\qquad\qquad
\rho_2(u)=\frac{dx}{du_2(x)}\Big|_{u_2(x)=u}.
\ee

To verify that this root configuration
is the $SO(6)$ singlet, we can take the large $u$
limit of \refb{contbe}
and assume that $\rho_1(u)$ and $\rho_2(u)$ fall off faster than
$u^{-2}$ as $u\to\infty$.  This shows that
\be
\int_{-\infty}^\infty du'\rho_1(u')=1\qquad\qquad
\int_{-\infty}^\infty du'\rho_1(u')=1/2,
\ee
which means that there are $L$ $u_1$ impurities and $L/2$ $u_2$ and
$u_3$ impurities, precisely what is needed to take the large $J$ state
to the singlet.

We can solve for $\rho_1(u)$ and $\rho_2(u)$ in \refb{contbe} by Fourier
transforming.  Defining
\be
\widetilde\rho_1(k)=\int du \exp(iku)\rho_1(u)\qquad\qquad
\widetilde\rho_2(k)=\int du \exp(iku)\rho_2(u),
\ee
it is straightforward to show that the solutions of \refb{contbe} are
\be
\widetilde\rho_1(k)=\frac{\cosh(k/2)}{\cosh(k)}\qquad\qquad
\widetilde\rho_2(k)=\frac{1}{2\cosh(k)},
\ee
Transforming back gives us
\be\label{rhosol}
\rho_1(u)=\frac{\cosh(u\pi/2)}{\sqrt{2}\cosh(u\pi)}\qquad\qquad
\rho_2(u)=\frac{1}{4\cosh(u\pi/2)}.
\ee
The anomalous dimension can now be computed and is
\be\label{anomdc}
\gamma=\frac{\lambda}{8\pi^2}E=\frac{\lambda}{8\pi^2}L\int_{-\infty}^\infty du
\frac{\rho_1(u)}{u^2+1/4}=\frac{\lambda}{8\pi^2}L\left(\frac{\pi}{2}+\ln2\right).
\ee

Not surprisingly, the anomalous dimension is extensive: it depends linearly on
$L$.  However, recall that in the BMN limit, we saw that two impurities
with the same real momentum had to have their roots split off from the
real line.  Hence if all the roots are real, each  $u_1$ impurity
has to correspond to a string oscillator with a different level number. 
Since there are $L$ such impurities and since they are equally distributed
between left and right oscillators, we find that the total level $\ell_{tot}$ 
is
\be
\ell_{tot}=\sum_{\ell=1}^{L/2}\ell\approx L^2/8.
\ee
Therefore, we see that the {\it full} 
dimension of the operator has the behavior
\be\label{levdim}
\Delta=L+\gamma=\sqrt{\ell_{tot}}\left(2\sqrt{2}+\frac{\lambda}{2\sqrt{2}\pi^2}
\left(\frac{\pi}{2}+\ln2\right)\right)+{\rm O}(\lambda^2),
\ee
the same square root dependence on the level that is generic for
small $\al'$ in string
theory \cite{9802109}.  
Of course, small $\al'$ corresponds to strong coupling where the
dimension of the operator had a 
$(\lambda)^{1/4}$ dependence.  In any event, \refb{levdim} suggests
that the square root dependence of the level is generic, even at weak 
coupling.  Note that corrections coming from higher orders in perturbation
theory should also give contributions to the dimension which are linear in
$L$, since the large $N$ expansion essentially localizes the interactions
to nearby neighbors.

\begin{figure}[h]
   %\hspace*{4cm}
   \begin{center}
   \epsfxsize=8cm
   \epsfbox{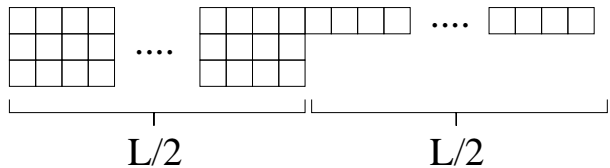}
\caption[x]{\small
Young tableau corresponding to the antiferromagnet configuration}
   \end{center}
   \label{young1}
   \end{figure}

Although the level square root dependence appears to be generic,
the actual $\lambda$ dependence depends on the operator under consideration.
For example, let us consider the  operator whose $SU(4)$ Young tableau is 
shown in figure \ref{young1}.
The corresponding Bethe state has $L/2$ $u_1$ excitatations and no $u_2$ and
$u_3$ excitations.  We then have the first equation in \refb{contbe} but with
$\rho_2(u)=0$.  This is same equation found for the anti-ferromagnetic Heisenberg
spin chain.  Its solution is well known (e.g. see \cite{9605187}).  
The anomalous dimension is
\be\label{anomhaf}
\gamma=\frac{\lambda}{8\pi^2}E=\frac{\lambda}{8\pi^2}L\int_{-\infty}^\infty du
\frac{\rho_1(u)}{u^2+1/4}=\frac{\lambda}{4\pi^2}L\ln2.
\ee
The anomalous dimension is smaller than in \refb{anomdc}, but so is the level,
since there are only $L/2$ excitations.  For this particular state, we
see that the full dimension is
\be\label{levdim2}
\Delta=L+\gamma=\sqrt{\ell_{tot}}\left(4\sqrt{2}+\frac{\lambda\sqrt{2}}{\pi^2}
\ln2\right)+{\rm O}(\lambda^2).
\ee
Thus, this has the level square root dependence, but the $\lambda$ 
dependence is different than that in \refb{levdim}. 

One can also consider ``excitations'' \cite{Resh3,dVK}
away from this $SO(6)$ singlet
by including ``holes''  in the integers appearing in \refb{ataneqs}.  The
inclusions of these holes modifies the 
equations in \refb{contbe} to
\ben\label{contbeh}
\frac{2}{4u^2+1}&=&\pi\rho_1(u)+\pi\sum_{j=1}^{\nt_1}\delta(u-\ut_{1,j})+
\int_{-\infty}^\infty du'
\frac{\rho_1(u')}{(u-u')^2+1}\nonumber\\
&&\qquad\qquad\qquad\qquad\qquad\qquad-2\int_{-\infty}^\infty du'
\frac{(\rho_2(u')+\rho_3(u'))}{4(u-u')^2+1}\nonumber\\
0&=&\pi\rho_2(u)+\pi\sum_{j=1}^{\nt_2}\delta(u-\ut_{2,j})+\int_{-\infty}^\infty du'
\frac{\rho_2(u')}{(u-u')^2+1}\nonumber\\
&&\qquad\qquad\qquad\qquad\qquad\qquad-2\int_{-\infty}^\infty du'
\frac{\rho_1(u')}{4(u-u')^2+1}\nonumber\\
0&=&\pi\rho_3(u)+\pi\sum_{j=1}^{\nt_3}\delta(u-\ut_{3,j})+\int_{-\infty}^\infty du'
\frac{\rho_2(u')}{(u-u')^2+1}\nonumber\\
&&\qquad\qquad\qquad\qquad\qquad\qquad-2\int_{-\infty}^\infty du'
\frac{\rho_1(u')}{4(u-u')^2+1},
\een
where $\nt_i$ refers to the number of holes of type $i$ and $\ut_{i,j}$ are
the positions of the holes.
Assuming that $\nt_i<<L$, the corrections from the $\delta$-functions to 
the densities are additive, so
we can consider them individually.  It is convenient to write the densities
as
\be
\rho_1(u)=\rho_1^{(0)}(u)+\frac{1}{L}\s_1(u)\qquad
\rho_2(u)=\rho_2^{(0)}(u)+\frac{1}{L}\s_2(u)\qquad
\rho_3(u)=\rho_3^{(0)}(u)+\frac{1}{L}\s_3(u)
\ee
where $\rho_i^{(0)}$ are the densities with no holes present.

For a hole of type 1 at position $\ut_1$,  we can write
\be
\s_1(u)=\s_1^1(u-\ut_1)\qquad\s_2(u)=\s_2^1(u-\ut_1)\qquad
\s_3(u)=\s_3^1(u-\ut_1).
\ee
The equations in \refb{contbeh} become
\ben\label{contbeh1}
0&=&\pi\s_1^1(u)+\pi\delta(u)+
\int_{-\infty}^\infty du'
\frac{\s_1^1(u')}{(u-u')^2+1}-4\int_{-\infty}^\infty du'
\frac{\s_2^1(u')}{4(u-u')^2+1}\nonumber\\
0&=&\pi\s_1^1(u)
\int_{-\infty}^\infty du'
\frac{\s_2^1(u')}{(u-u')^2+1}-2\int_{-\infty}^\infty du'
\frac{\s_1^1(u')}{4(u-u')^2+1}\nonumber\\
\een
where we have used the symmetry of the configuration to set $\s_2^1(u)=\s_3^1(u)$.  The equations in \refb{contbeh1} are easily solved, giving
\ben
\s_1^1(u)&=&-\int \frac{dk}{2\pi}e^{-iku}\frac{e^{-|k|/2}\cosh(k/2)}{\cosh(k)}
\nonumber\\
\s_2^1(u)=\s_3^1(u)&=&-\int \frac{dk}{2\pi}e^{-iku}\frac{e^{-|k|/2}}{2\cosh(k)}.
\een
The change in the energy is
\be\label{ehole}
\eps(\ut_1)=\int_{-\infty}^\infty du \frac{\s_1(u-\ut_1)}{u^2+1/4}
=-2\pi\rho_1^{(0)}(\ut_1),
\ee
where $\rho_1^{(0)}(u)$ is the solution in \refb{rhosol}.

To find the momentum of the hole, we can integrate $\eps(\ut_1)$ with respect
to $\ut_1$, giving
\be
p(\ut_1)=\pi-2\arctan\left(\sqrt{2}\sinh\frac{\ut_1\pi}{2}\right).
\ee
Notice that $Lp(\ut_1)/(2\pi)$ is the change in the level coming from
the introduction of the hole, which can be easily deduced by looking at
\refb{ataneqs}, \refb{contbe} and \refb{ehole}.
It is also possible  to express $\eps$ in terms of $p$, where we find
\be
\eps(p)=-\pi\sin\left(\frac{p}{2}\right)
\sqrt{1+\sin^2\left(\frac{p}{2}\right)}\qquad\qquad 0\le p<2\pi.
\ee
and so the change in the anomalous dimension is
\be
\Delta\gamma=-\frac{\lambda}{8\pi}\sin\left(\frac{p}{2}\right)
\sqrt{1+\sin^2\left(\frac{p}{2}\right)}
\ee

In order to understand the nature of these holes, notice that
\be
\int du\s_1^1(u)=1, \qquad \int du\s_2^1(u)= \int du\s_3^1(u)
=\frac{1}{2}.
\ee
Hence, we need an even number of these types of holes\footnote{We can have
an odd number, but we need to add another lattice site.}.  We also see that
the highest weight of each hole is $\vec w=\vec\al_1 +\frac{1}{2}(\vec\al_2+\vec\al_3)$
which is the highest weight of the  $SO(6)$ vector representation. Hence
these holes come with a vector index.

Next consider a  type 2 hole.  Proceeding as before, we find that
\ben
\s_1^2(u)&=&-\int \frac{dk}{2\pi}e^{-iku}\frac{e^{-|k|/2}}{1+e^{-2|k|}}
\nonumber\\
\s_2^2(u)&=&-\int \frac{dk}{2\pi}e^{-iku}
\frac{1+e^{-|k|}+e^{-2|k|}}{(1+e^{-|k|})(1+e^{-2|k|})}\nonumber\\
\s_3^2(u)&=&-\int \frac{dk}{2\pi}e^{-iku}
\frac{e^{-|k|}}{(1+e^{-|k|})(1+e^{-2|k|})}.
\een
The energy of this type of hole is
\be
\eps(\ut_2)=\int du\frac{\s_1^2(u-\ut_2)}{u^2+1/4}=-2\pi\rho_2^{(0)}(\ut_2),
\ee
where $\rho_2^{(0)}(\ut_2)$ is the density in \refb{rhosol}.  Integrating
$\eps$, we find that the momentum is
\be
p(\ut_2)=\pi-2\arctan(e^{\pi\ut/2}),
\ee
and so the energy of this type of hole in terms of $p$ is
\be
\eps(p)=-\frac{\pi}{2}\sin p\qquad 0\le p\le\pi.
\ee
Hence these holes occupy only half of a Brillouin zone.
We also have that 
\be\label{notype2}
\int du\s_1^2(u)=\frac{1}{2}, \qquad \int du\s_2^2(u)= \frac{3}{4}
\qquad\int du\s_3^2(u)=\frac{1}{4},
\ee
thus the highest weight of each type 2 hole is
$\vec w=\frac{1}{2}\vec\al_1+\frac{3}{4}\vec\al_2+\frac{1}{4}\vec\al_3$
which is the highest weight of one of the spinor representations.  

The argument for type 3 holes is the same as for type 2.  The highest weight
of each type 3 hole is $\vec w=\frac{1}{2}\vec\al_1+\frac{1}{4}\vec\al_2+\frac{3}{4}\vec\al_3$, hence each of these type holes is in the other spinor
representation.  Since the two spinor representations are complex conjugates,
we choose the energies of the type 3 holes to be
\be
\eps(p)=+\frac{\pi}{2}\sin p\qquad \pi\le p\le 2\pi.
\ee
 
The trace condition forces the total momentum of the holes to be zero
mod $2\pi$.
We
can also see from \refb{notype2} that every type 2 hole has to either 
come with three other type 2 holes, or a type 3 hole.  The same is true
for type 3 holes.  These conditions tell us that we cannot have individual
spinor excitations, since the chain itself is made up of $SO(6)$ vectors.
Instead the representations have to combine to form an adjoint rep, or
another representation that is trivial under the $SO(6)$ center.\footnote{If
$L$ is odd, then the excitations combine to form a vector representation,
or another representation which has the same action under the 
center of $SO(6)$.}

\sectiono{Conclusions}

In this paper we constructed a mixing operator for anomalous dimensions and
showed that it was related to the Hamiltonian of an integrable $SO(6)$ chain.
We then used the Bethe ansatz to find the anomalous dimensions of many operators, including those that were outside  the BMN limit.  We also demonstrated
that these non-BMN operators have anomalous dimensions that depend on
the square root of the level, a result also found at strong coupling.

There are many other operators where it is hoped that the Bethe ansatz
will allow one to compute anomalous dimensions.  These include the operators
that correspond to large wound strings oscillating on the $S^5$.  A prediction
was made for the anomalous dimensions based
on a semiclassical analysis \cite{0209047}, 
and it would be nice to explicitly verify this.

It would also be nice if one could somehow relate 
the higher loop corrections to
integrable Hamiltonians.  One possibility is that the higher loop corrections
correspond to the higher Hamiltonians in the heirarchy of the same spin chain.
On one level, this seems reasonable.  In the large $N$ limit, one would expect
the $g$ loop corrections to the anomalous dimensions to involve mixing between
$g+1$ nearest neighbors, which is precisely what is found in the $g^{\rm th}$
Hamiltonian in the heirarchy.  
However,  this idea does not appear to work.  
For example, the two-loop analysis as done in
\cite{0205066} shows that the two-loop anomalous dimension matrix should
have operators of the form
\be
\sum_l^L(P_{l,l+1}P_{l+1,l+2}+P_{l+1,l+2}P_{l,l+1}).
\ee
But the next Hamiltonian in the hierarchy 
of a Heisenberg system has the form
\be
\sum_l^L(iP_{l,l+1}P_{l+1,l+2}-iP_{l+1,l+2}P_{l,l+1}).
\ee

That this idea does not work is perhaps not too surprising.
 If the higher Hamiltonians of the hierarchy were indeed 
related to anomalous
dimensions at higher orders of perturbation theory, then
 the mixing matrix could have been
diagonalized by a unitary transformation which is independent of the 
coupling --- a rather exceptional property.
In any event, one can now ask what role the higher hamiltonians play
on the gauge theory side.
  
\bigskip
\noindent {\bf Acknowledgments}:
We are grateful to A.~Tseytlin and P.~Wiegmann  for discussions.
K.Z. would like to thank the Erwin Schr\"odinger Institute and 
J.A.M. would like to thank the CTP at MIT for
hospitality provided during the course of this work.
This research was
supported in part by the Swedish Research Council.
The work of K.Z. was also supported in part by
RFBR grant  01-01-00549 and in part by  grant
00-15-96557 for the promotion of scientific schools.

\end{document}